\documentstyle[aps,prl,epsfig,floats]{revtex}
\begin{document}
\draft
\narrowtext
\twocolumn
\wideabs{
\title{Prospects for gravitational-wave observations of neutron-star tidal disruption in neutron-star/black-hole binaries}
\author{Michele Vallisneri\cite{email}}
\address{Theoretical Astrophysics 130-33, California Institute of Technology, Pasadena CA 91125. \\
INFN, Sez.\ di Milano, Gruppo Collegato di Parma/Universit\`a di Parma, 43100 Parma, Italy.}
\date{December 7, 1999}
\maketitle
\begin{abstract}
For an inspiraling neutron-star/black-hole binary (NS/BH), we estimate the gravity-wave frequency $f_{\rm td}$ at the onset of NS tidal disruption. We model the NS as a tidally distorted, homogeneous, Newtonian ellipsoid on a circular, equatorial geodesic around a Kerr BH.
We find that $f_{\rm td}$ depends strongly on the NS radius $R$, and estimate that LIGO-II (ca.~2006--2008) might measure $R$ to 15\% precision at $140$ Mpc ($\sim 1$ event/yr under current estimates).
This suggests that LIGO-II might extract valuable information about the NS equation of state from tidal-disruption waves.
\end{abstract}
}
\noindent The equation of state of the bulk nuclear matter inside a neutron star (NS) is poorly understood \cite{Baym92}. For example, candidate equations of state that are compatible with nuclear physics experiments and theory predict, for a $1.4 \, M_\odot$ NS, a radius anywhere from about $8 \, {\rm km}$ to $16 \, {\rm km}$ \cite{Cutler90}.
Thorne has conjectured that insights into the equation of state might come from measurements of the gravitational waveforms emitted by merging NS/NS binaries and/or tidally disrupting NS's in neutron-star/black-hole (NS/BH) binaries \cite{Thorne87,Abramovici92}. More recently, Newtonian models of NS/NS mergers have given strong evidence that the merger waves {\em do} carry equation-of-state information, but for NS/NS are emitted at frequencies ($\sim$ 1400--2800 Hz) too high for measurement by LIGO-type gravity-wave interferometers \cite{Zhuge94,Hughes98}. In this paper, we show that the prospects for NS/BH measurements are much brighter.

\begin{figure}
\begin{center}
\epsfig{file=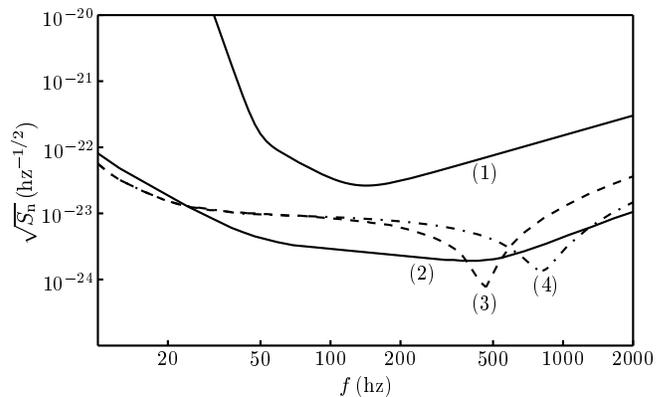, width=3.4in}
\caption{Plot of (square root of) noise spectral density for different LIGO configurations: (1) LIGO-I; (2) LIGO-II wideband; (3), (4) LIGO-II narrowband centered on 500 and 850 Hz. Curves (1)-(3) are from \protect\cite{Lsc99}; curve (4) was produced by K.\ A.\ Strain using the same detector specifications as in \protect\cite{Lsc99}.
\label{fig:ligos}}
\end{center}
\end{figure}

Central to these prospects is the question of whether NS tidal-disruption waves lie in the band of good interferometer sensitivity (for LIGO-II, $\sim$ 30--1000 Hz \cite{Lsc99}; see Fig.~\ref{fig:ligos}).  Numerical modeling of NS tidal disruption in NS/BH binaries is only now getting underway \cite{Lee98} and has not yet included computations of the emitted gravity waves or even their frequency bands. As a result, the best frameworks now available for estimating the tidal-disruption gravity-wave band are highly simplified, quasi-analytic models by Shibata \cite{Shibata96} and by Wiggins and Lai \cite{Wiggins99}, which represent the inspiraling NS as an irrotational \cite{Bildsten92}, incompressible or polytropic Newtonian ellipsoid, moving on a circular, equatorial geodesic orbit around a Kerr BH, and being tidally distorted by the Kerr Riemann tensor.  For simplicity we focus on Shibata's homogeneous models, and then appeal to the polytropic models for evidence that compressibility has only small effects.

In Shibata's analysis, the NS gravitational field, its centrifugal potential, and the Newtonian tidal potential constructed from the Kerr Riemann tensor are all quadratic functions of position. As a result, a class of equilibrium solutions are the classic irrotational, homogeneous {\em Roche-Riemann ellipsoids} \cite{Chandrasekhar69}. Given a choice of the binary parameters $M$, $a$ and $r$ (the BH mass and angular momentum per unit mass, and the orbital separation, i.~e., the Boyer-Lindquist radius of the geodesic), there is a one-parameter family of such NS models with density $\rho$ ranging downward through the family to a minimum $\rho_{\rm cr}(M, a, r)$.

We model the inspiraling NS as one of Shibata's irrotational ellipsoids, identified by its mass $m$, and its density $\rho$ or mean radius $R = (3 m / 4 \pi \rho)^{1/3}$. In our simple framework, the uncertainty in $m(R)$ embodies the uncertainty about the NS equation of state. We describe the inspiral as a sequence of circular, equatorial Kerr geodesics that shrink inward until the NS reaches the innermost stable circular orbit, $r=r_{\rm isco}$, or begins to tidally disrupt [which happens at the radius $r_{\rm td}$ where the star's density $\rho$ matches the critical density $\rho_{\rm cr}(M, a, r_{\rm td})$].

The Kerr geometry provides a one-to-one correspondence between the orbital radius $r_{\rm td}$ and the gravity-wave frequency $f_{\rm td}$ at which tidal disruption begins:
\begin{equation}
\label{eq:freqtoradius}
f_{\rm td}(M,a,r_{\rm td}) = \frac{1}{\pi(a + \sqrt{r_{\rm td}^3 / M)}}
\end{equation}
(here and below we set $G = c = 1$). It is this $f_{\rm td}$ that LIGO-II can measure.  Having measured $f_{\rm td}$ and determined the masses $M$ and $m$ from the observed inspiral waveforms \cite{Cutler94}, one can compute $r_{\rm td}$ and then the NS density $\rho = \rho_{\rm cr}(M,a,r_{\rm td})$ and the mean NS radius $R$. Thereby, the LIGO-II observations can determine a point on the NS mass-radius curve $m(R)$, which represents the NS equation of state in our simplified analysis.  Even one such point could give valuable information about the {\em real} NS equation of state, and several such points could determine it remarkably well \cite{Lindblom92}.

\begin{figure}
\begin{center}
\epsfig{file=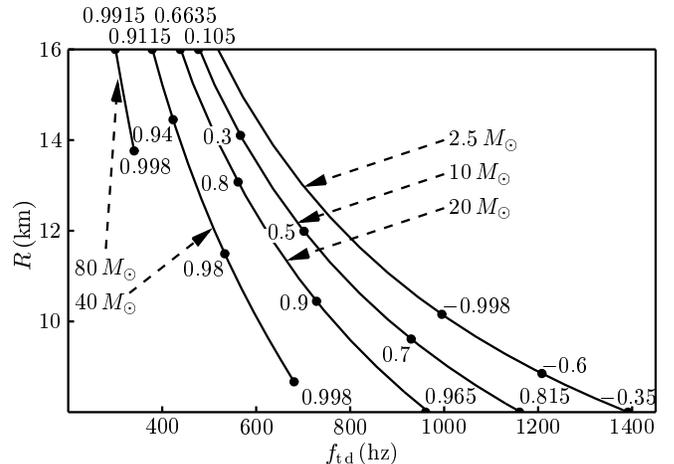, width=3.4in}
\caption{
NS radius $R$ vs.\ disruption-onset frequency $f_{\rm td}$, for $m = 1.4 \, M_\odot$ and $M=2.5$--$80 \, M_\odot$.  The black dots, parametrized by $a/M$, specify the onset of plunge into the BH; tidal disruption is measurable only for $f_{\rm td}$ left of the plunge point, i.~e., for $R$ above it. Negative $a/M$ indicates retrograde NS orbits.
\label{fig:holes}}
\end{center}
\end{figure}

To estimate the accuracy with which LIGO-II might determine the NS radius $R$, we need the explicit relationship between $R$ and the disruption-onset frequency $f_{\rm td}$. More precisely, we need $R(m,M,a,f_{\rm td})$, which can be derived as follows: (i) $r_{\rm td}(M,a,f_{\rm td})$ is obtained by inverting Eq.~(\ref{eq:freqtoradius}); (ii) $\rho_{\rm cr}(M,a,r_{\rm td})$ is obtained by solving Eq.~(3.9) of \cite{Shibata96} for the ratios of semiaxes of the equilibrium configurations, and then extremizing Eq.~(3.10) of \cite{Shibata96}, in which $\tilde{\Omega}^2 = M/(\pi \rho r^3)$; (iii) then $R$ is obtained as $R=[3 m / 4 \pi \rho_{\rm cr}(M,a,r_{\rm td})]^{1/3}$. The result has the form
\begin{equation}
\label{eq:nsradius}
R(m,M,a,f_{\rm td}) = m^{1/3} M^{2/3} \, \hat{D} \Bigl[ \frac{a}{M}, f_{\rm td} M \Bigr],
\end{equation}
where $\hat{D}$ is a dimensionless function with remarkably weak dependence on $a/M$ \cite{foot2}.  This $R(f_{\rm td})$ is shown in Fig.~\ref{fig:holes} for various $M$, for $a/M=0.998$ (the curves for other $a/M$ are almost identical to these), and for $m = 1.4 \, M_\odot$ \cite{Thorsett99}. The radii shown, $R = 8 \mbox{--} 16 \, {\rm km}$ for $m = 1.4 \, M_\odot$, correspond to the range of predictions by plausible NS equations of state \cite{Cutler90}.  The curves in Fig.\ \ref{fig:holes} are well approximated by the formula (with $G=c=1$)
\begin{equation}
\label{eq:approximations}
\frac{R}{m^{1/3} M^{2/3}} \approx
\Biggl\{
\begin{array}{ll}
0.145 \, (f_{\rm td} M)^{-0.71} & \text{for $f_{\rm td} M \lesssim 0.045$}, \\
0.069 \, (f_{\rm td} M)^{-0.95} & \text{for $f_{\rm td} M \gtrsim 0.045$}.
\end{array}
\end{equation}

Although the BH spin parameter $a$ has negligible influence on the function $R(f_{\rm td})$, it strongly influences the radius $r_{\rm isco}$ of the innermost stable circular orbit \cite{foot1b}. If the NS is still intact when it reaches $r_{\rm isco}$, it then will plunge rapidly into the BH and the tidal-disruption waves, if any, are likely to be so weak and short-lived as to be useless for measuring NS properties. Thus, there is not much hope of measuring tidal disruption unless $f_{\rm td} < f_{\rm plunge} = [$Eq.\ (\ref{eq:freqtoradius}) with $r_{\rm td}$ replaced by $r_{\rm isco}(M,a)$]; i.~e., unless $f_{\rm td}$ is left of the relevant big dot in Fig.~\ref{fig:holes}. 

Figure \ref{fig:holes} and the above discussion show that (i) for a wide range of realistic parameters, tidal disruption occurs before the plunge begins, and (ii) for all realistic parameters except a very narrow range ($M \alt 10 \, M_\odot$ and $R \alt 10 \, {\rm km}$), the tidal-disruption waves fall in the range of good LIGO sensitivity, $f \alt 1000 \, {\rm Hz}$.  The Lai-Wiggins polytropic NS models \cite{Wiggins99} give similar curves and conclusions: for polytropic indices $n=0.5$ and $1.0$, which approximate NS equations of state, the $R(f_{\rm td})$ curves are displaced upward in frequency from those of Fig.\  \ref{fig:holes} by a mere $\sim 50$ and $\sim 100 \, {\rm Hz}$. 

Turn now to an estimate of the accuracy to which LIGO-II could measure $f_{\rm td}$ (and then $R$) using {\em Wiener optimal filtering} \cite{Wainstein62,Finn92}. The measured gravity-wave data stream $g(t)$ is compared  to a set of theoretical inspiral {\em templates} $h(\theta^i;t)$, indexed by the parameters $\theta^i$ of the binary; a ``best fit'' $\hat{\theta}^i$ is found which maximizes the likelihood of observing $g(t)$ given a ``true'' signal $h(\hat{\theta}^i;t)$, and given a statistical model of the detector noise [a Gaussian \cite{foot3} random process with zero mean and spectral density $S_n(f)$].
For strong enough signals, $\hat{\theta}^i$ will have a gaussian distribution centered around its ``true value'' $\tilde{\theta}^i$, with covariance matrix \cite{Finn92}
\begin{equation}
\label{eq:covariance}
C_{ij} = (\Gamma^{-1})_{ij}, \; \; \Gamma_{ij} \approx 2 \left\langle \frac{\partial h}{\partial \theta^i}(\hat{\theta}^k) \, \right| \left. \frac{\partial h}{\partial \theta^j}(\hat{\theta}^k) \right\rangle,
\end{equation}
where the ``inner product'' $\langle \ldots \rangle$ is defined for any two real data streams $g(t)$, $h(t)$ in terms of their Fourier transforms $\tilde{g}(f)$, $\tilde{h}(f)$ by
\begin{equation}
\label{eq:innerproduct}
\langle g , h \rangle = \int_{-\infty}^{\infty} df \,
\frac{\tilde{g}(f) \tilde{h}^*(f)}{S_n(|f|)}.
\end{equation}

Because so little is known about the tidal disruption and our NS models are so crude, we use the simplest of templates in our analysis: slow-motion, quadrupolar waveforms for point particles in circular, Keplerian orbits with quadrupole-governed inspiral. The Fourier-transformed waveform, squared and averaged over binary directions and orientations, is given by \cite{foot3ab}
\begin{equation}
\label{eq:squaredstrain}
\langle |\tilde{h}_b|^2 \rangle = \frac{\pi}{30}
 \frac{\mu M_T^3}{d^2} \frac{1}{(\pi M_T f)^{7/3}} \, \theta(f_{\rm plunge} - f),
\end{equation}
where $\mu$ and $M_T$ are the reduced and total masses, $d$ is the distance to the binary, and the step function shuts off the signal at the onset of plunge.

For typical observations, optimal filtering of the inspiral signal should give good estimates of $M$ and $m$ \cite{Cutler94}. We therefore assume that the accuracy in measuring $R$ is limited only by the uncertainty of $f_{\rm td}$ \cite{foot3b}.
The estimation of $f_{\rm td}$ depends heavily on the details of the tidal-disruption waveforms, which are largely unknown.  However, it is reasonable to expect tidal disruption to be a sudden event that significantly weakens gravity-wave emission within a few dynamical time-scales after $f_{\rm td}$ has been reached \cite{foot3c}. Correspondingly, we employ a toy model where the inspiral waveform of Eq.~(\ref{eq:squaredstrain}) dies out over a frequency band $(f_{\rm td},f_{\rm td}+\delta f)$:
\begin{equation}
\label{eq:tdwave}
\tilde{h}_{\rm td}(f) =
\left\{
\begin{array}{ll}
\tilde{h}_b(f) &
\text{if $f < f_{\rm td}$}, \\
\tilde{h}_b(f) \Theta(\frac{f - f_{\rm td}}{\delta f}) &
\text{if $f_{\rm td} < f < f_{\rm td} + \delta f$}, \\
0 &
\text{if $f > f_{\rm td} + \delta f$},
\end{array}
\right.
\end{equation}
where $\Theta(x) = 1 - x$ (linear decay), or $\Theta(x) = 10^{-x}$ (exponential decay).
The standard deviation of the ``best fit'' $\hat{f}_{\rm td}$ is given by Eq.~(\ref{eq:covariance}) as $\Delta \hat f_{\rm td} = [\Gamma_{f_{\rm td}f_{\rm td}}(\tilde{h}_{\rm td})]^{-1/2}$.

We have evaluated $\Delta \hat f_{\rm td}$ numerically, using the signal model from Eqs.~(\ref{eq:squaredstrain}), (\ref{eq:tdwave}) and the inner product (\ref{eq:innerproduct}) with the LIGO-II noise curves $S_n(f)$ of Fig.~\ref{fig:ligos}.
We have then computed the $2 \sigma$ range of the NS radii $R$ from the relation $R_\pm = R(m,M,a,\hat{f}_{\rm td} \mp 2 \Delta \hat{f}_{\rm td})$ [Eq.~(\ref{eq:nsradius})]. The uncertainty in $R$, defined as $\Delta R = (R_+ - R_-)/2$, scales roughly linearly with $d$ [because $\Delta \hat{f}_{\rm td}$ is proportional to $d$ through Eqs.~(\ref{eq:covariance}), (\ref{eq:squaredstrain})], and is quite sensitive to the choice of the shutoff model [it scales roughly as $(\delta f)^{1/2}$ and is lower for the exponential decay than for the linear one].
In Table \ref{tab:errors} we report the fractional uncertainty $\Delta R/R$, averaged over the range $10 \, {\rm km} < R < 15 \, {\rm km}$, for choices of parameters motivated by the following.

The NS mass $m$ was set to be $1.4 \, M_\odot$ \cite{Thorsett99}. The distance $d$ and the BH masses $M$ were chosen to represent two different scenarios:
(i) low-mass BH's, with $M = 2.5 \, M_\odot$ at $65 \, {\rm Mpc}$ ($\sim$ one merger/yr according to Bethe and Brown \cite{Bethe98});
(ii) higher mass BH's, with $M = 10$, $20$, and $40 \, M_\odot$, at $140 \, {\rm Mpc}$ (massive main-sequence binaries are thought to produce NS/BH binaries with $M \sim 10 \, M_\odot$ and coalescence rates up to $\sim$ one event/yr out to $140 \, {\rm Mpc}$, but possibly much less \cite{bhrates}; capture NS/BH binaries formed in globular clusters might have $M$ as large as hundreds of $M_\odot$ \cite{Kulkarni93}, but with exceedingly uncertain rates).
Finally, we considered three different gravity-wave shutoff models:
(i) an optimal-precision model with linear decay and $\delta f = f_{\rm td}/6$ (the lower limit set by the uncertainty principle on the frequency spread of waves emitted during 3 orbital periods, supposedly a typical time-scale for complete disruption \cite{foot3c}); %
(ii) a fiducial model with exponential decay and $\delta f = f_{\rm td}/2$ (a scaling supported by numerical calculations of tidal-disruption waveforms in NS/NS binaries \cite{Zhuge94}); this model was also used to evaluate errors for $n=1$ polytropes;
(iii) a conservative model with linear decay and $\delta f = f_{\rm td}/2$.
\begin{table}
\caption{
Fractional uncertainty $\Delta R/R$, averaged over the range $10 \, {\rm km} < R < 15 \, {\rm km}$. Rows: BH masses; columns: gravity-wave decay models and detector noise curves [labeled by (2)--(4) as in Fig.~\protect\ref{fig:ligos}]. No quote is given if $\Delta R/R > 25$\%.\label{tab:errors}}
\begin{tabular}{lcccccccccccccccc}
& & \multicolumn{15}{c}{$\Delta R/R$ (\%) for $10 \, {\rm km} < R < 15 \, {\rm km}$} \\
\tableline
& &
\multicolumn{3}{c}{lin.\ decay} & &
\multicolumn{3}{c}{exp.\ decay} & &
\multicolumn{3}{c}{e.\ d., $n = 1$} & &
\multicolumn{3}{c}{lin.\ decay} \\
& &
\multicolumn{3}{c}{$\delta f = f_{\rm td}/6$} & &
\multicolumn{3}{c}{$\delta f = f_{\rm td}/2$} & &
\multicolumn{3}{c}{$\delta f = f_{\rm td}/2$} & &
\multicolumn{3}{c}{$\delta f = f_{\rm td}/2$} \\
\tableline
$M \, \backslash \, S_n$ & &
(2) & (3) & (4) & &
(2) & (3) & (4) & &
(2) & (3) & (4) & &
(2) & (3) & (4) \\
\tableline
$2.5 \, M_\odot$\tablenotemark[1] & &
12 & 18\tablenotemark[3] & 8 &&
20 & -- & 13 &&
24\tablenotemark[3] & -- & 25 &&
21\tablenotemark[3] & -- & 17 \\
$10 \, M_\odot$\tablenotemark[2] & &
14 & 17\tablenotemark[3] & 10 &&
23 & -- & 17 &&
25\tablenotemark[3] & -- & 21 &&
25\tablenotemark[3] & -- & 19 \\
$20 \, M_\odot$\tablenotemark[2] & &
10 & 16 & 10 &&
16 & 14\tablenotemark[3] & 16 &&
23 & -- & 14 &&
22 & 25\tablenotemark[3] & 16 \\
$40 \, M_\odot$\tablenotemark[2] & &
7 & 6 & 11 &&
11 & 10 & 19 &&
20 & 10\tablenotemark[3] & 20 &&
17 & 19 & 23
\end{tabular}
\tablenotemark[1]{At 65 Mpc.}
\tablenotemark[2]{At 140 Mpc.}
\tablenotemark[3]{For $12 \, {\rm km} < R < 15 \, {\rm km}.$}
\end{table}

The estimates for our fiducial decay model suggest that $R$ may be determined with a precision of $\sim$ 15\% using the $850 \, {\rm Hz}$-narrowband LIGO-II configuration [curve (4) of Fig.~\ref{fig:ligos}], and with a somewhat worse precision for wideband LIGO-II [curve (2)]. If the optimal-precision decay model is correct, the error might be as low as $\sim$ 6--10\%. The usefulness of the $500 \, {\rm Hz}$-narrowband interferometer [curve (3) of Fig.~\ref{fig:ligos}] is limited to the the heavier BH's or to the larger NS's, which have lower $f_{\rm td}$. Our estimates are inferior for the Lai-Wiggins compressible polytropes \cite{Wiggins99} examined in the least favorable case ($n=1$), and for the most conservative decay model; even then, an $850 \, {\rm Hz}$-narrowband LIGO-II might be able to provide significant information about $R$.

The accuracy of our analysis is limited by several factors. Sources of error in the frequency $f_{\rm td}(m,M,a,R)$ at which tidal disruption begins to significantly change the inspiral waveforms include:
(i) the use of the test-mass approximation for the NS orbit, when actually $m \not\ll M$, especially for the low-mass Bethe-Brown case;
(ii) the use of the Riemann tensor to compute tidal forces when the NS diameter is not, typically, small compared to the distance from the NS center to the horizon \cite{foot5};
(iii) the idealization of the NS as a homogeneous or polytropic ellipsoid;
(iv) the fact that the point at which the observed waveforms show a clear deviation from a standard inspiral may actually come a few orbits earlier (due to tidal coupling) or later than $f_{\rm td}$.

Our method presupposes a reliable technique to distinguish a plunge shutoff of the inspiral waves from a tidal-disruption shutoff.  In fact, it seems likely that the tidal-disruption waveform will actually contain features that not only distinguish it from a plunge shutoff, but that also carry equation-of-state information which is richer than in our crude model. For example, simulations \cite{Zhuge94} of tidal disruption in NS/NS binaries show a spectrum with an inspiral cutoff followed by a valley, a moderately sharp peak, and a cliff; however, the NS/BH case is likely to be different, and the issue will ultimately be settled only by detailed numerical simulations.

Given these large uncertainties, our results can only be rough indications of the prospects for learning about NS's from tidal-disruption waveforms.  They do, however, suggest that observations of tidal disruption in NS/BH binaries might be possible in $\sim$ 2006--2008 with LIGO-II, and might yield useful insights into the NS equation of state. The success of this endeavor will require the development of better theoretical and numerical techniques for modeling NS tidal disruption and computing the dependence of the disruption waveforms on the NS equation of state; we strongly advocate such an effort.

The author thanks Kip Thorne for insight, inspiration and support; for useful interactions, thanks go to M.\ Pauri, B.\ Allen, K.\ A.\ Strain, V.\ Kalogera, L.\ Lindblom, C.\ Cutler, S.\ Gon\c{c}alves, J.\ Creighton, T.\ Creighton, and K.\ Alvi. This research was supported in part by NSF Grants AST-9731698 and PHY-9900776.

\begin{thebibliography}{10}
%
\bibitem[*]{email} E-mail: vallis@tapir.caltech.edu
\bibitem{Baym92}
G. Baym, in {\it Isolated Pulsars}, proceedings of the Los Alamos Workshop, Taos, New Mexico, 1992, edited by K.~A.~Van Riper, R.~Epstein, and C.~Ho (Cambridge University Press, 1993), p.\ 1.
\bibitem{Cutler90}
C. Cutler, L. Lindblom, and R.~J. Splinter, Astrophys. J. {\bf 363}, 603 (1990).
\bibitem{Thorne87}
K.~S. Thorne,  in {\em Three hundred years of gravitation}, edited by S.
Hawking and W. Israel (Cambridge University Press, 1987), p.\ 330.
\bibitem{Abramovici92}
A. Abramovici {\em et al.}, Science {\bf 256}, 325 (1992).
\bibitem{Zhuge94}
X. Zhuge, J.~M. Centrella, and S.~L.~W. McMillan, Phys. Rev. D {\bf 50},  6247 (1994); {\bf 54}, 7261 (1996).
\bibitem{Hughes98}
S. Hughes, Ph.D. thesis, Caltech, 1998;
F.~A. Rasio and S.~L. Shapiro, Class. Quantum Grav. {\bf 16}, R1 (1999).
\bibitem{Lsc99}
E. Gustafson, D. Shoemaker, K. Strain, and R. Weiss, {\em LSC White Paper on Detector Research and Development} (LIGO-Project document, September 11, 1999).
\bibitem{Lee98}
See Wiggins and Lai \cite{Wiggins99} for references.
\bibitem{Shibata96}
M. Shibata, Prog. Theor. Phys. {\bf 96},  917  (1996); for earlier tidally-locked models see L.~G. Fishbone, Astrophys. J. {\bf 175},  L155  (1972); {\bf 185}, 43 (1973); {\bf 195}, 499  (1975).
\bibitem{Wiggins99}
P.~Wiggins and D.~Lai, Astrophys.\ J. {\bf 532}, 530 (2000); see also D.~Lai, F.~Rasio and S.~L.~Shapiro, {\it ibid.} {\bf 423}, 344 (1994).
\bibitem{Bildsten92}
L. Bildsten and C. Cutler, Astrophys. J. {\bf 400}, 175 (1992) have shown that the NS cannot be tidally locked to the BH, so Shibata's irrotational models are more realistic than Fishbone's tidally locked ones \cite{Shibata96}.
\bibitem{Chandrasekhar69}
S. Chandrasekhar, {\em Ellipsoidal figures of equilibrium} (Yale University Press, New Haven, 1969).
\bibitem{Cutler94}
C. Cutler and {\`E}.~E. Flanagan, Phys. Rev. D {\bf 49},  2658  (1994).
\bibitem{Lindblom92}
L. Lindblom, Astrophys. J. {\bf 398},  569  (1992).
\bibitem{foot2}
This is because apart from a weak dependence on $a/M$, the orbital frequency of Kerr geodesics scales with orbital separation in the same way as the tidal strength, $\sim M / r^3$.
\bibitem{Thorsett99}
$1.4 \, M_\odot$ is the value, to within $\sim \pm 0.1 \, M_\odot$, of all well-measured NS masses in NS/NS binaries. See, e.~g., S.~E. Thorsett and D. Chakrabarty, Astrophys. J. {\bf 512}, 288 (1999).
\bibitem{foot1b}
J.~M. Bardeen, W.~H. Press, and S.~A. Teukolsky, Astrophys. J. {\bf 178}, 347 (1972).
\bibitem{Wainstein62}
L.~A. Wainstein and L.~D. Zubakov, {\em Extraction of signals from noise} (Prentice-Hall, Englewood Cliffs, 1962).
\bibitem{Finn92}
L.~S. Finn, Phys. Rev. D {\bf 46},  5236  (1992).
\bibitem{foot3}
It is expected that non-Gaussian noise will be removed by coincidence between several detectors.
\bibitem{foot3ab}
See Eq.~(44) of \cite{Thorne87}, but change $\pi/12$ to $\pi/6$ (typo). A further factor of $1/5$ accounts for detector orientations.
\bibitem{foot3b}
For a signal to noise ratio $\agt 10$ (fairly typical of the observations examined in this paper), and if spins can be treated as negligible, $\Delta m / m$, $\Delta M / M \alt 0.02$ \cite{Cutler94}, and from Eq.~(\ref{eq:approximations}) the influence of $\Delta m$ and $\Delta M$ on $\Delta R$ gives $\Delta R / R \sim 0.005$. If spins are important these errors increase tenfold, but might be considerably reduced if the {\em a priori} knowledge of $m$ from known NS/NS binaries \cite{Thorsett99} can be applied to NS/BH systems.
\bibitem{foot3c}
Bildsten and Cutler \cite{Bildsten92} estimate that complete disruption would take place in $\sim$ 1--3 orbital periods, while the disrupted NS would spread into a ring in $\sim$ 1--2 periods, significantly reducing the gravity-wave amplitude. These rough estimates are confirmed qualitatively by numerical simulations of NS/NS binaries (see \cite{Bildsten92} for references).
\bibitem{Bethe98}
H.~A. Bethe and G.~E. Brown, Astrophys. J. {\bf 506}, 780 (1998).
\bibitem{bhrates}
V. Kalogera, in Proceedings of the 3rd Amaldi Conference on Gravitational Waves, Caltech, 1999 (to be published, astro-ph/9911532), and references therein. Extrapolation of event rates from the Galaxy to the Universe follows S.~E. Phinney, Astrophys. J. {\bf 380}, L17 (1991).
\bibitem{Kulkarni93}
S.~R. Kulkarni, P. Hut and S. McMillan, Nature {\bf 364}, 421 (1993);
S. Sigurdsson and L. Hernquist, {\em ibid.} {\bf 364}, 423.
\bibitem{foot5}
The ratio $r_{\rm td} / 2 R$ is $\sim$ 3--5 for $m = 1.4 \, M_\odot$ and $M = 2.5 \, M_\odot$; $\sim 5$ for $M = 10 \, M_\odot$; $\sim$ 6--8 for $M = 20 \, M_\odot$. The ratios are slightly higher if we use the proper distance to the BH horizon instead of $r_{\rm td}$.
%
\end{thebibliography}
\end{document}